\documentclass[prl, twocolumn,superscriptaddress]{revtex4}
\usepackage{amsmath,amssymb,bm}
\usepackage{graphicx}
\usepackage{epstopdf}
\usepackage{latexsym}
\usepackage{subfigure}
\usepackage{color}
\usepackage{natbib}
\usepackage{braket}
      
\begin{document}

\title{SO($N$) singlet projection model on the kagome lattice}

\author{Matthew S. Block}
\affiliation{Department of Physics \& Astronomy, California State University, Sacramento, CA 95819}

\begin{abstract}
  We present an extensive quantum Monte Carlo study of a nearest-neighbor, singlet-projection model on the kagome lattice that exhibits SO($N$) symmetry and is sign-problem-free. We find that in contrast to the previously studied SU($N$) variations of this model, the non-bipartite lattices appear to harbor spin-liquid phases for intermediate values of $N$, a result also seen on the triangular lattice with this same model. Unlike the triangular lattice, however, the kagome lattice appears to admit no valence bond solid (VBS) phase for large $N$, instead remaining a spin-liquid.  We also observe that the spin-ordered phase survives to a relatively large value of $N$, at least $N=8$, and that it is gone for $N=10$; the fate of $N=9$ remains unclear.
\end{abstract}
\date{\today}
\maketitle

\section{Introduction}
The search for exotic quantum phase transitions in two dimensions (2D) has been a fruitful endeavor from the perspective of numerical investigations. Of note is the prediction, detection, and characterization of so-called deconfined quantum critical points. These critical points defy traditional Landau-Ginzburg-Wilson theory of phase transitions by allowing for a direct, continuous transition between two phases that break fundamentally different symmetries. Furthermore, numerical evidence corroborated the claim that at the critical point, an emergent U(1) gauge field mediates interactions between spinon degrees of freedom that are normally confined in the adjacent phases.\cite{senthil2004:science} The numerical linchpin to the success of these studies was the development of SU($N$)-symmetric spin-singlet projection models deployed on several different bipartite 2D lattices.\cite{sandvik2007:deconf,lou2009:sun,kaul2012:j1j2} A natural extension is to consider the same type of sign-problem-free operator on a non-bipartite lattice, such as the kagome, where the symmetry is merely SO($N$). That is precisely what we endeavor to do here.

\section{Model}
We consider the kagome lattice where each site has a Hilbert space of $N$ states, denoted for site $j$ as $\Ket{\alpha}_j$ where $\alpha=1,\ldots,N$. By using the fundamental representation of SO($N$) on each site it is possible to construct spin singlets on any two sites: $\Ket{S_{ij}}=\frac{1}{\sqrt{N}}\sum_\alpha\Ket{\alpha\alpha}_{ij}$. We can then construct the singlet-projection operator for a pair of sites as $\hat{\mathcal{P}}_{ij}=\Ket{S_{ij}}\Bra{S_{ij}}$.  This follows closely the previous numerical study on the triangular lattice~\cite{kaulTriangular}. We consider this operator acting on nearest neighbors of the lattice and this is the first term in our model Hamiltonian:
\begin{equation}
\label{eqn:J1term}
\hat{\mathcal{H}}_{J_1}=-J_1\sum_{\braket{ij}}\hat{\mathcal{P}}_{ij}.
\end{equation}
We can study this model on its own for integer values of $N$ to map out the phase diagram as a function of the symmetry order (see Results below). To gain a more detailed understanding of the phase transition between observed phases, we can add a second term that acts on the shortest bonds joining sites on the same sublattice (denoted $\{ij\}$; see Fig.~\ref{fig:kagome}):
\begin{figure}[t]
\centerline{\includegraphics[width=\columnwidth]{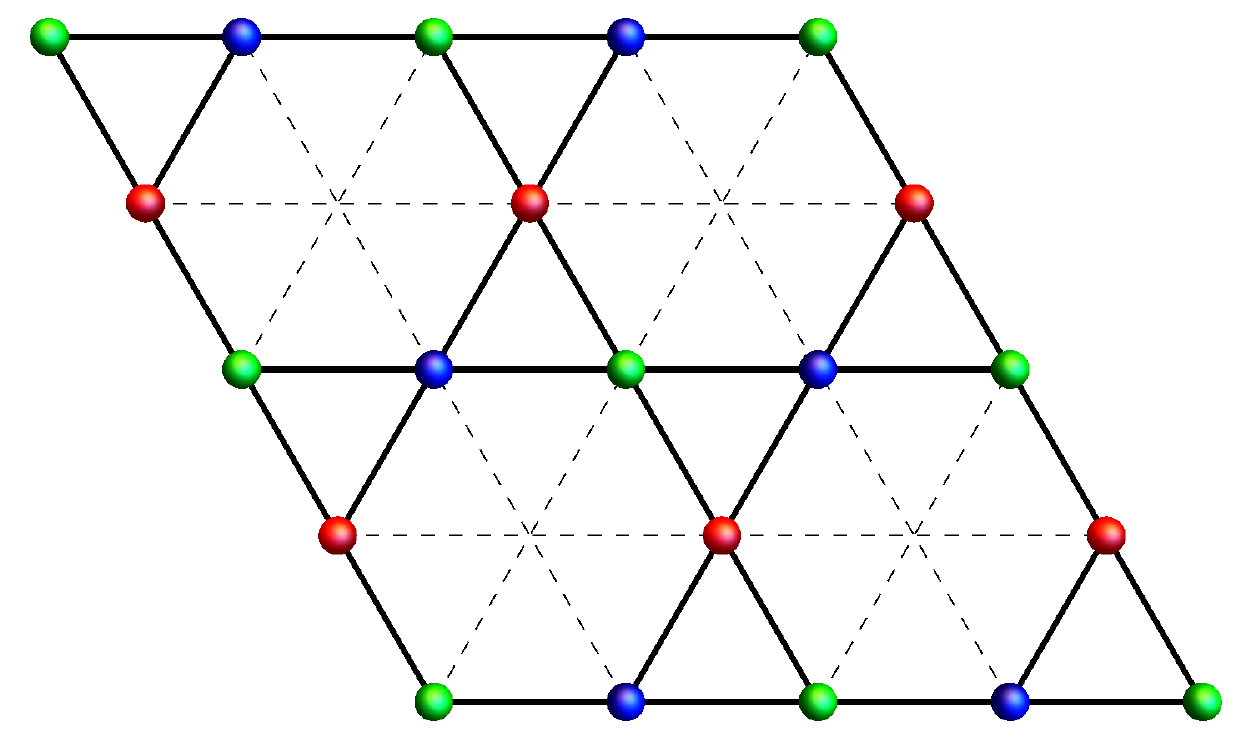}}
\caption{(color online).  A small cluster of the kagome lattice. The solid lines connect nearest-neighbor sites where the coupling is $J_1$ while the dotted lines connect nearest-neighbor sites \emph{on the same sublattice} where the coupling is $J_2$.}
\label{fig:kagome}
\end{figure}

\begin{equation}
\label{eqn:J2term}
\hat{\mathcal{H}}_{J_2}=-J_2\sum_{\{ij\}}\hat{\Pi}_{ij},
\end{equation}
where $\hat{\Pi}_{ij}=(1/N)\sum_{\alpha,\beta}\Ket{\alpha\beta}_{ij}\Bra{\beta\alpha}_{ij}$, the so-called permutation operator, which encourages spin ordering.  Our full model is thus $\hat{\mathcal{H}}=\hat{\mathcal{H}}_{J_1}+\hat{\mathcal{H}}_{J_2}$.
By varying $g\equiv J_2/J_1$, we can continuously tune from the spin-liquid phase for some large $N$ to the spin-ordered phase and perform a detailed study of the properties of the phase transition. A detailed study along these lines will be presented in a later publication.

\section{Measurements}
In all cases, we study lattices with $N_\text{site}=L_x\times L_y$ sites with $L\equiv L_x=L_y$.  The ``$x$" and ``$y$" axes are chosen to be along straight lines that connect lattice sites 120$^\circ$ apart. Periodic boundary conditions are enforced along these directions, which preserves the rotational symmetry of the lattice. We employ the stochastic series expansion method for our quantum Monte Carlo, which samples via local bond updates and global loop updates. Aside from some proprietary measurement code, the QMC algorithm was developed and described in detail by Anders Sandvik.~\cite{sandvik2010:vietri}

We will now discuss the particular numerical measurements we make for this study.  First, to assess the spin-ordered phase, we introduce the $N\times N$ spin operator:
\begin{equation}
\label{eqn:Qab}
\hat{Q}_{\alpha\beta}=\Ket{\alpha}\Bra{\beta}-\frac{\delta_{\alpha\beta}}{N}
\end{equation}
and the corresponding spin-spin correlation function:
\begin{equation}
\label{eqn:Cspin}
C_{\text{spin}}(\mathbf{r}_i-\mathbf{r}_j)=\sum_\alpha\Braket{\hat{Q}_{\alpha\alpha}(i)\hat{Q}_{\alpha\alpha}(j)}.
\end{equation}
The spatial Fourier transform gives the spin static structure factor:
\begin{equation}
\label{eqn:sssf}
S(\mathbf{k})=\frac{1}{N_\text{site}}\sum_\mathbf{r}e^{i\mathbf{k}\cdot\mathbf{r}}C_\text{spin}(\mathbf{r}),
\end{equation}
which will exhibit a Bragg peak at $\mathbf{k}=\mathbf{0}$ when spin order is present, the height of which will diverge as the system size is increased.  Values of the spin static structure factor at neighboring values of $\mathbf{k}$ will correspondingly die off.  To facilitate detection of the spin order, we can therefore define a spin order parameter as follows:
\begin{equation}
\label{eqn:sop}
R_\text{S}=1-\frac{S_\text{spin}(\text{peak neighbor})}{S_\text{spin}(\text{peak})},
\end{equation}
which will trend to zero if no spin order is present and will trend to one when spin order is present as $L\rightarrow\infty$.

Next, to search for VBS order, we first define a simple bond operator intended to correspond to every nearest-neighbor bond on the lattice indexed by $b=1,\ldots,N_\text{bond}$:
\begin{equation}
\label{eqn:Pb}
\hat{\mathcal{P}}_b(\tau)=\left(\Ket{S_{b[1]b[2]}}\Bra{S_{b[1]b[2]}}\right)_\tau,
\end{equation}
where $b[1]$ and $b[2]$ are site indices referring to the first and second sites, respectively, on a given bond. Here, the label $\tau$ indicates a moment in discretized imaginary time within our Monte Carlo scheme (more details below). The corresponding VBS correlation function with appropriate subtraction is:
\begin{equation}
\label{eqn:Cvbs}
C_{\text{VBS}}(\mathbf{r}_i-\mathbf{r}_j,\tau)=\Braket{\hat{P}_{b_i}(\tau)\hat{P}_{b_j}(0)}-\Braket{\hat{P}_{b_i}(\tau)}\Braket{\hat{P}_{b_j}(0)}.
\end{equation}
Now we perform a spacetime Fourier transform at zero-frequency to obtain the VBS susceptibility:
\begin{equation}
\label{eqn:VBSsus}
\chi_\text{VBS}(\mathbf{k})=\frac{1}{N_\text{site}}\sum_\mathbf{r}e^{i\mathbf{k}\cdot\mathbf{r}}\frac{1}{\beta}\int d\tau \, C_\text{VBS}(\mathbf{r},\tau),
\end{equation}
where $\beta$ is the usual reciprocal temperature. Why we use imaginary-time-dependent operators for some measurements and not for others is purely based on efficiency of the measurement within our particular Monte Carlo scheme; these details do not affect the conclusions.  
%
%Finally, as with the spin order, we define a VBS order parameter to facilitate detection of the phase:
%\begin{equation}
%\label{eqn:vbsop}
%R_\text{VBS}=1-\frac{\chi_\text{VBS}(\text{peak neighbor})}{\chi_\text{VBS}(\text{peak})}.
%\end{equation}
%
A critical difference here from the spin case is that we do not know the location of the Bragg peaks, if any, at the outset.  Instead, we must search for peaks and assess their scaling as $L\rightarrow\infty$.

\section{Results}
\begin{figure}[t]
\centerline{\includegraphics[width=\columnwidth]{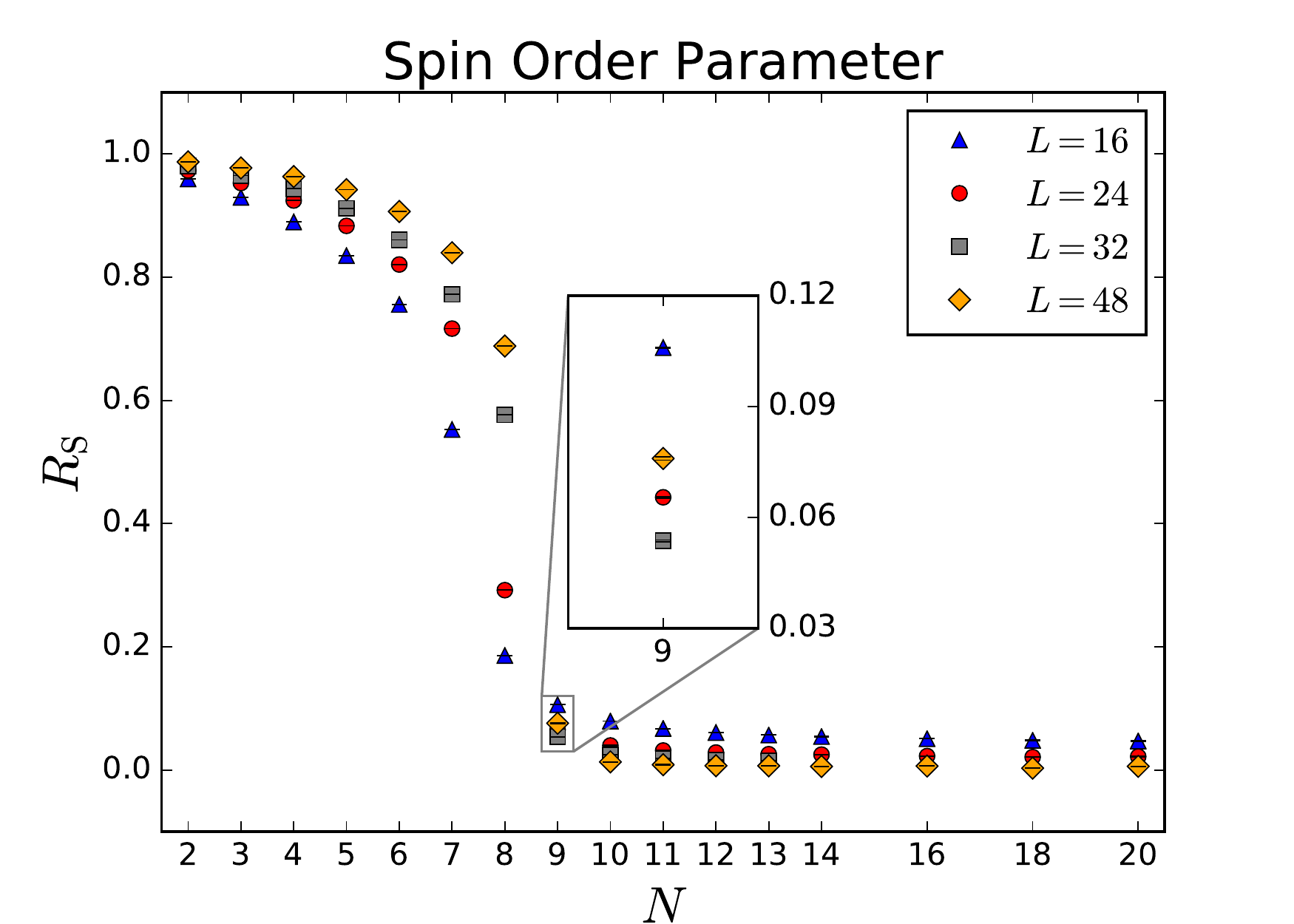}}
\caption{(color online).  The spin order parameter, $R_\text{S}$ as a function of $N$.  A value of 1 indicates spin-order. All values of $N$ show the quantity approaching either 1 or 0 monotonically with increasing $N$ except for $N=9$.  The inset shows the non-monotonicity.}
\label{fig:RSvsN}
\end{figure}

We began by studying the $J_1$-only model while varying $N$ to determine how the symmetry order affected the realized phase.  A modest investigation using system sizes $L=16, 24, 32, 48$ (Fig.~\ref{fig:RSvsN}) revealed that spin order persists for $N\leq8$ and is definitely not present for $N\geq10$. At $N=9$, there is a very weak signal in the spin order parameter, but the data shows a non-monotonicity suggesting that for larger system sizes, the signal might strengthen and spin order may indeed be present. It was not deemed a worthwhile expenditure of resources to find out since the spin-ordered phase could be well understood from the smaller values of $N$ and, should the spin order fail to persist, the resulting phase would most likely be the same as for the larger values of $N$, a spin liquid.

In past investigations, such as the one referenced earlier on the triangular lattice, the spin-liquid phase eventually gave way to a VBS phase for large $N$ and there exist dimer models in the large-$N$ limit that provide reasonable explanations for this behavior. In the case of the kagome lattice, however, the question of what should theoretically happen at large $N$ remains murky. We had hoped to provide evidence that would aid in settling this question, but, regrettably, our code has significant limitations in the non-spin-ordered regime and especially for values of $N$ approaching 20.  If and when a VBS pattern with a reasonably small unit cell exists, we can detect it reliably. This was the case even for the relatively extended $\sqrt{12}\times\sqrt{12}$ VBS present for this model on the triangular lattice for large $N$.~\cite{kaulTriangular,moessner2001:isingfrus,ralko:dimertriang} However, confirming the absence of a VBS phase is significantly more challenging as a weak signal could be difficult, if not impossible, to discern from the noise. Error bars for the VBS order parameter tend to be large and the large system sizes necessary to realize an extended and potentially incommensurate VBS pattern quickly become intractable.  Examination of $\chi_\text{VBS}(\mathbf{k})$ in search of potential Bragg peaks is hindered by these issues, but no prominent or consistent peak locations were evident for large values of $N$ across a range of system sizes.  For this reason, we feel comfortable concluding that no VBS order is present, at least up through $N=20$.

\section{Conclusion}
Our quantum Monte Carlo study of the SO($N$) projector model on the kagome lattice reveals the existence of only two detectable phases: a spin-ordered phase for small values of $N$ and a spin-liquid phase showing no signs of ordering in the ground state for larger $N$. This result is in contrast to the study of the same model on the triangular lattice, which, though it found an intervening spin-liquid phase, concluded that the spin-liquid phase eventually gave way to VBS order for sufficiently large $N$. We propose further study of the large-$N$ scenario on the kagome lattice to verify the absence of a VBS phase. A field-theoretic prediction of what a stable VBS phase would look like on the kagome or evidence that no such phase should form would aid enormously in this effort. On the numerical front, we can make the $J_1$ coupling stronger along one natural axis of the kagome lattice to force VBS order. By analyzing the pattern, we can identify where the Bragg peaks would occur in momentum space were VBS order to manifest in the isotropic case.  This would allow for a more controlled search for VBS order. 

Preliminary investigations of the phase transition between the spin-ordered and spin-liquid phase (using the $J_1$-$J_2$ model described above) suggest a continuous transition with somewhat bizarre critical behavior.  A more detailed study of this transition is warranted.

There exists ample opportunity for extensions of these investigations.  In particular, we plan to consider our same model on the three-dimensional pyrochlore lattice.  To our knowledge, a controlled numerical study of the SO($N$) projector model on a three-dimensional lattice has not been undertaken. Should a spin-liquid also be realized there, it will be significant to catalog how the critical properties of the transition to the neighboring spin-ordered phase measure against the already catalogued two-dimensional examples.

\section{Acknowledgements}
We gratefully acknowledge the National Science Foundation and, in particular, the XSEDE collaboration and the San Diego Supercomputer Center whose Comet cluster was instrumental to this study. The research reported here was supported in part by NSF DMR-130040.

\end{document}